\documentclass[showpacs,prl,letterpaper,floatfix,nobalancelastpage,twocolumn]{revtex4}
\usepackage{amsmath}
\usepackage{bm}
\usepackage{graphicx}
\usepackage{color}
\usepackage{float}
\usepackage{graphicx,subfigure}

\def\ket#1{\mathinner{|{#1}\rangle}}
\def\braket#1{\mathinner{\langle{#1}\rangle}}

\begin{document}

\title{Stationary inversion of a two level system coupled to an off-resonant cavity\\ with strong dissipation}
\author{S. Hughes$^{1,2}$}
\email{shughes@physics.queensu.ca}%
\author{H. J. Carmichael$^{2}$}
\affiliation{$^1$Department of Physics, Queen's University, Kingston, Ontario, Canada K7L 3N6}
\affiliation{$^2$Department of Physics, University of Auckland, Private Bag 92019, Auckland, New Zealand}

\begin{abstract}
We present an off-resonant excitation scheme that realizes pronounced stationary inversion in a two level system. The created inversion exploits a cavity-assisted two photon resonance to enhance the multi-photon regime of nonlinear cavity QED and survives even in a semiconductor environment, where the cavity decay rate is comparable to the cavity-dot coupling rate. Exciton populations of greater than 0.75 are obtained in the presence of realistic decay and pure dephasing. Quantum trajectory simulations and quantum master equation calculations help elucidate the underlying physics and delineate the limitations of a simplified rate equation model.
Experimental signatures of inversion and multi-photon cavity QED 
are predicted in the  fluorescence intensity 
and second-order correlation function measured as a function of drive power.
\end{abstract}

\maketitle

The term  ``population inversion'' refers to a system in which an excited state population exceeds that of a lower level. Its use is of broad interest to laser science. Semiclassical arguments readily show that a steady state drive, e.g., a continuous wave (cw) laser, cannot create population inversion in a two level system (TLS) because of stimulated emission; thus, standard atomic and molecular lasers typically exploit a cascaded through three, four or more levels to achieve inversion \cite{Eberly:book}. In 1988, an important limitation of the semiclassical argument was pointed out by Savage \cite{Savage:PRL88}, who demonstrated that the multi-quanta regime of an atom-based cavity QED system can achieve {\em stationary} population inversion in a TLS. Specifically, by including the effects of {\em quantum noise}, and with the pump resonant with the degenerate atom-cavity resonance, Savage \cite{Savage:PRL88}---and later Lindberg and Savage  \cite{LindbergSavage:PRA88}---predicted achievable inversions of around 0.01-0.03. The unanticipated result was explained in terms of a 4-state two-quanta model of an on-resonance  cavity-atom system in the bad-to-intermediate cavity regime. Lindberg and Savage \cite{LindbergSavage:PRA88} suggest that the inversion might be measured in high pump resonance fluorescence, as a small excess (a few $\%$) over the saturation fluorescence intensity. To the best of our knowledge this prediction has not been verified by experiment on either atomic or semiconductor systems.

There have been numerous recent experiments in semiconductor cavity systems
using coherent excitation, most exploiting a deliberate detuning between the
cavity and quantum dote (QD). For example, Majumdar {\em et al.}~\cite{Majumdar:PRB10} studied the coupling between a photonic crystal cavity and an off-resonant QD under coherent excitation of either the cavity or the QD; they investigated exciton power broadening under coherent excitation in the presence of a detuned cavity. Ulhaq {\em et al.} \cite{Ultaq:PRB10} investigated linewidth broadening of a resonantly excited QD in a micropillar cavity, again with QD-cavity detuning. Both groups demonstrate significant power broadening of the target QD exciton, and, importantly, the experimental capacity to explore off-resonant QD-cavity systems using nonlinear coherent excitation in a controlled way. There have also been remarkable developments in circuit QED \cite{FinkNature:2008,Bishop:Nature09,JohnsonNature:2010}, where anharmonic cavity QED effects are realized at dipole coupling strengths two to three orders of magnitude larger than in atomic or QD systems.

In this Letter, we introduce a scheme to create
pronounced
population inversion in a dissipative TLS coupled to a off-resonant cavity. Supported by quantum trajectory (QT) simulations \cite{Tian:PRA92,Dalibard:PRL92} and quantum master calculations, we show that {\em off-resonant} excitation of a detuned TLS-cavity system, with intermediate to strong coupling, results in population inversions that can easily exceed 0.25. These stationary inversions are enabled through two-photon excitation of the dressed states, and, strategically, {\em only} occur for a detuned TLS and cavity. In fact, significant prototype inversions have been seen in a circuit QED device excited via so-called {\em sideband transitions}~\cite{Leek:PRB09}, an excitation scheme similar to ours; however, the unanticipated inversion is discussed only briefly in terms of a simplified rate equation model. We use QT simulations to, first, connect to, then extrapolate beyond rate equations. We show that they have a limited range of validity, and focus on the possibility of achieving significant inversion far outside the rate equation regime. We show that such inversion exists for practical semiconductor systems where it is potentially important for microlasers and few photon quantum light sources.
Other non-standard routes to gain include lasing without inversion~\cite{Carmichael:PRA97}, but here
we are dealing with conventional inversion-driven gain, but with an unfamiliar pumping scheme.
We also suggest experiments for measuring the inversion, via the fluorescence intensity measured as a function of drive power \cite{Majumdar:PRB10,Ultaq:PRB10}, or through the second-order correlation function; both exhibit clear signatures of the proposed off-resonant stationary inversion regime.

\begin{figure}[t!]
\includegraphics[width=8.4cm]{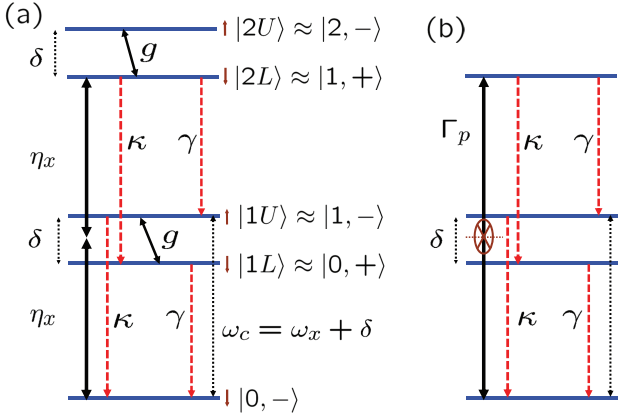}
\vspace{-0.0cm}
\caption{ (Color online) (a) Quasienergy levels truncated to 5 states ($\delta \equiv \delta_{cx} = \omega_c-\omega_x$). $\ket{1,+}$ refers to one created cavity photon with one exciton (excited TLS) created; for a far-detuned cavity and exciton, this doubly-excited state relates to the Jaynes-Cummings dressed states through $\ket{2L} \approx  \ket{1,+}$. Bold arrows illustrate two-photon resonance, with the pump detuned from the cavity by about $-\delta/2$. Small arrows show the direction of the Stark shifts. (b) Simplified rate equation model, where $\Gamma_p \equiv\Omega^2_{tp}/2\kappa$ is an effective two-photon pump rate, with $\Omega_{tp}$ the two-photon Rabi frequency (see text).
\label{fig1}}
\end{figure}

The cavity QED system is described through a quantum master equation for the reduced density matrix of the TLS (exciton in a semiconductor dot) and cavity. Working in the interaction picture, i.e., in a frame rotating at the pump frequency, $ \omega_L$,  the master equation is \cite{CarmichaelBook1}
\begin{eqnarray}
\frac{d\rho}{dt}=\frac{-i}{\hbar}[H_s, \rho]+\left(\kappa{\cal L}[a]+\frac\gamma2{\cal L}[\sigma_-]+\frac{\gamma^\prime}8{\cal L}[\sigma_z]\right)\rho,
\label{eqn:ME}
\end{eqnarray}
with system Hamiltonian
\begin{eqnarray}
 H_s&=&\hbar \delta_{xL} {\sigma}^+{\sigma}^-+\hbar \delta_{cL} {a}^\dagger{a} +\hbar g({\sigma}^-a^\dagger + \sigma^+a)\nonumber\\&& + \hbar \eta_x(\sigma^+ + \sigma^-),
 \label{eqn:Hamiltonian}
\end{eqnarray}
where $\eta_x$ is the coherent pump rate of the TLS, $g$ is the TLS-cavity coupling strength, $a$ is the cavity mode annihilation operator, ${\sigma}^+,{\sigma}^-$ and $\sigma_z$ are Pauli operators, and $\delta_{\alpha L}=\omega_\alpha-\omega_L$ are pump laser detunings; dissipation is described by the Lindblad ${\cal L[\xi]\rho}  = 2 \xi \rho\xi^\dagger -\xi^\dagger\xi\rho - \rho \xi^\dagger\xi$, with separate terms to account for decay of the cavity  mode (rate $\kappa$), and radiative decay (rate $\gamma$) and pure dephasing (rate $\gamma^\prime$) of the TLS.

Figure \ref{fig1}(a) depicts the dressed quasienergy levels of Hamiltonian (\ref{eqn:Hamiltonian}), truncated to 5 states. The lowest 4 levels are the main ones responsible for cavity-assisted inversion, where decay of $\ket{1,+}$ populates $\ket{0,+}$ through the emission of a cavity photon. Importantly, two photon resonance to  $\ket{1,+}$, with negligible excitation of $\ket{1,-}$, is achieved by setting $2\delta_{Lx}\approx \delta\equiv \delta_{cx}=\omega_c-\omega_x\gg g,\kappa,\gamma,\gamma^\prime$. For increasing pump strengths, Stark shifts lower the resonance frequency. For comparison, a simplified ``rate equation'' model, with adiabatic elimination of intermediate states, is shown in Fig.~\ref{fig1}(b).

We first derive the TLS population within the simplified model. With coherent drive, and neglecting pure dephasing, we obtain the steady-state population of the TLS excited state (population of excitons),
\begin{equation}
\bar n_x \equiv  \braket{\sigma^+\sigma^-}_{ss}= \frac{4\Omega_{tp}^2}{(\gamma+4\Omega_{tp}^2/2\kappa)(\gamma+2\kappa)},
\label{eqn:rate}
\end{equation}
with two-photon drive $\Omega_{tp} = (4\eta_x^2/\delta)^2/(g/\delta)^2$. Note that this expression differs from the one used by Leek {\em et al.}~\cite{Leek:PRB09}; their formula, $\bar n_x =\Omega_{tp}/(\Omega_{tp}+\gamma/2\kappa+\gamma\Omega_{tp}/2\kappa)$, is obtained by treating $\Omega_{tp}$ as the rate of an {\em incoherent} drive.

\begin{figure}[t!]
\centering
\subfigure{
\includegraphics[trim = 2.5mm 1mm 0mm 0mm, clip, width=8.6cm,height=7.2cm]{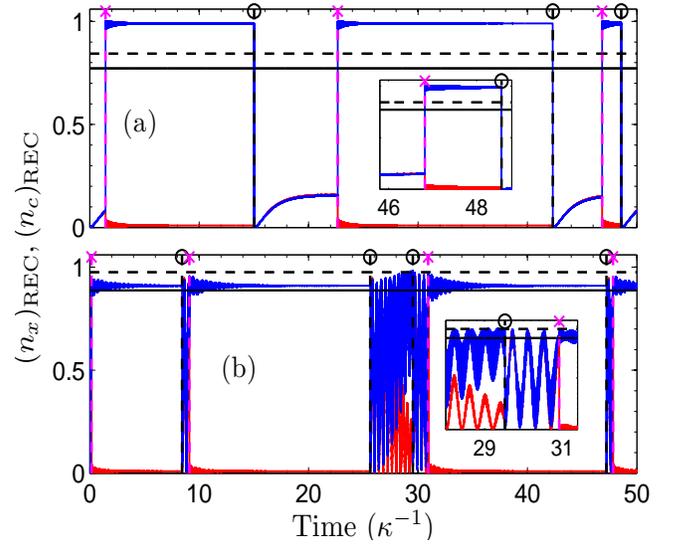}
}
\caption{ (Color online)
Sample QTs near, (a) [$g=1000\kappa$, $\eta_x=100\kappa$, $\gamma=0.05\kappa$], and outside, (b) [$g=100\kappa$, $\eta_x=150\kappa$, $\gamma=0.05\kappa$], the rate equation regime. The blue solid curve plots the expectation of the TLS excited population (exciton population) conditioned upon the record of quantum jumps, while the red solid curve is the conditioned expectation of the cavity photon number. Quantum jumps are indicated by the crosses and circles (and vertical dashed lines) for photon emission via the cavity and TLS, respectively. The steady-state expectation of the exciton population is shown with the horizontal solid line, while the rate equation estimate is shown with the dashed horizontal line. The insets enlarge the regions near $t\kappa \approx 46-48$ and $t\kappa \approx 29-31$ in (a) and (b), respectively. \label{fig2}
}
\end{figure}

Using QT simulations we identify a regime where the rate equation picture is approximately valid while also allowing for significant inversions. Broadly, we require: $\delta/\kappa \geq10^4$, $g/\kappa\geq 10^3$, and $\eta_x/\kappa\geq10^2$.
Figure~\ref{fig2}(a) shows a sample QT on the boundary of this region, with $\gamma/\kappa=0.05$. Taking level shifts into account, the two-photon resonance is near $\delta_{xL}=-0.49g$. We highlight the following: ($i$) the TLS jumps
to an excitation close to unity with the cavity excitation close to zero; ($ii$) each such jump is initiated by the emission of a photon by the cavity corresponding to the transition $\ket{1,+}\rightarrow\ket{0,+}$ in Fig.~\ref{fig1}(a); ($iii$) all excitation periods terminate with the emission of a photon by the TLS corresponding to the transition $\ket{0,+}\rightarrow\ket{0,-}$ in Fig.~\ref{fig1}(a); ($iv$) the intervals between excitation periods (e.g., $t\kappa\approx 15-22$) see the cavity and TLS populations grow continuously and in step, corresponding to coherent excitation of the $\ket{0,-}\rightarrow\ket{1,+}$ transition. Solid and dashed horizontal lines show that there is nevertheless a discrepancy between the exact (solving the ME in a suitably large photon number basis~\cite{Numerical:Note}) population ($\bar n_x\sim 0.77$) and that predicted by Eq.~(\ref{eqn:rate}) ($\bar n_x \sim 0.84$); most likely this is due to our neglect of level shifts when deriving $\Omega_{tp}$.

We next choose parameters that start to deviate away from the rate equation picture. Figure \ref{fig2}(b) is plotted for $\delta/\kappa=10^3$, $g/\kappa=10^2$, and $\eta_x/\kappa=1.5\times10^2$, still with $\gamma/\kappa=0.05$. This places us close to the circuit QED regime of Leek {\em et al.}~\cite{Leek:PRB09}. Although a similar jump pattern is seen, the following differences are noticed: ($i$) the jump ``up'' reaches an excitation probability significantly less than unity (a small piece of $\ket{1,U}\approx \ket{1,-}$ is mixed into the state realized after the jump); ($ii$) the drive strength has been increased, which induces Rabi oscillations on the $\ket{0,-}\rightarrow\ket{1,+}$ transition (e.g., right before $t\kappa=31$);
($iii$) the TLS and cavity do not continue indefinitely to Rabi oscillate in step, as for $t\kappa\approx26-26$, where the TLS shows a damped evolution toward a number near 0.9, while the cavity photon number decays to zero. This behavior illustrates a marked difference between QTs and rate-equation-inspired quantum jumps. The final state of the damped Rabi oscillation is
$\ket{1,L} \approx \ket{0,+}$; but this state is reached via a coherent evolution in response to the back action of a null measurement---the system localizes to this state because no photon happened to be scattered through the cavity; the two-photon resonance still assists passage to the inverted state $\ket{1,L}\approx \ket{0,+}$, but through the absence rather than the presence of a quantum jump.

QT simulations have clarified two keys points: (a) the rate equation model is extremely restricted even for circuit QED parameters, and (b) significant population inversion can be achieved when dissipation rates increase relative to the dipole coupling. This suggests that significant inversions might be achieved using semiconductor QD parameters.

Turning now to
semiconductor QED systems, we  adopt the following scaled parameters in the strong coupling regime: $\kappa=g/2.5$, $\gamma=g/200$, and $\gamma'=g/25$. For current strongly coupled QD systems, $g$ is around $0.1\,$meV ($\approx 24\,$GHz), with similar values measured in a wide range of devices, including examples comprised of  photonic crystals \cite{SC:StrongCoupling1}, micropillars \cite{SC:StrongCoupling2}, and microdisk cavities \cite{SC:StrongCoupling3}. To connect with measurements of intensity linewidth broadening (e.g., Refs.~\cite{Majumdar:PRB10,Ultaq:PRB10}), we calculate the intensity of the exciton mode, $I_x \propto \bar n_x$, or the cavity mode, $I_c \propto \bar n_c$. We solve the full ME, Eq.~(\ref{eqn:ME}), in a basis that can be truncated at an arbitrary photon/exciton state, allowing us to compute both weak-excitation-approximation results (truncation at one quantum) and the regime of multi-quanta cavity QED. While weak excitation is commonly assumed when analyzing dissipative semiconductor cavity systems, we will show that the approximation drastically breaks down in the proposed excitation regime.

\begin{figure}[t]
\centering
\subfigure{
\includegraphics[trim = 3.2mm 0mm 0mm 0mm, clip, width=8.6cm,height=7.2cm]{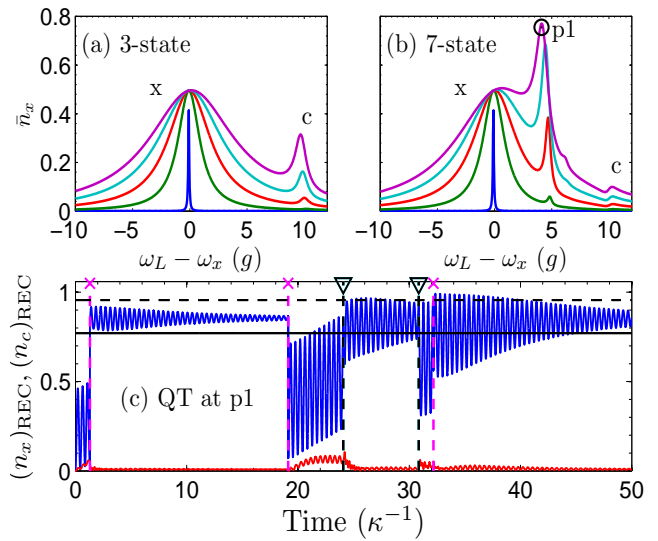}
}
\caption{ (Color online)
{\em Semiconductor quantum-dot QED system.}
(a-b) Pump-dependent steady-state exciton number versus detuning calculated with 1-photon and 3-photon truncations; for $\eta_x=0.02g-2g$ (blue to purple) and $\delta=10g$. Note the clear peak in the 3-photon case near $\omega_L-\omega_x\approx\delta/2$. (c) Sample QT simulation on the peak labeled p1; symbols are defined as in Fig.~2, with the additional pure dephasing {\em phase-glitch} indicated by an inverted triangle.\label{fig3}}
\end{figure}

We display in Figs.~\ref{fig3}(a-b) the exciton population as a function of laser-exciton detuning, with $\delta=10\,g$,  for two different truncations. All calculations predict substantial power broadening and show the expected peaks at the bare cavity and exciton resonances, in agreement with recent weak-coupling experiments \cite{Majumdar:PRB10,Ultaq:PRB10}. With multi-photon effects included, however, and for sufficiently large drive, the two-photon resonance emerges approximately half way between $\omega_x$ and $\omega_c$. The resonance yields pronounced exciton populations, exceeding $n_x=0.75$ for $\eta_x \approx 2g$ (inversions of $w=n_x-0.5>0.25)$. The position of this intermediate ``resonance'' changes with the pump strength, which gives a direct signature of the Stark-shifted two photon resonance. Similar peaks appear in the cavity intensity (not shown). As a strong contrast, on-resonance excitation ($\delta=0$) yields only very small inversions (a few \%) \cite{Savage:PRL88,LindbergSavage:PRA88}.

What is surprising about these predictions is that they show a pronounced influence from higher lying ladder states, which are notoriously difficult to see in a semiconductor system because of the relatively large cavity broadening. Recent experiments exploring QD anharmonicities  have been reported by Kasprzak {\em et al.}~\cite{Kasprzak:Nature10}, using advanced coherent nonlinear spectroscopy. The signatures observed are very weak. On the other hand, the scenario we investigate shows a profound impact from the second rung of the Jaynes-Cummings ladder, specifically the state $\ket{2-}\approx\ket{1,+}$, where a quantized level of the radiation field serves as a rung in a multi-level pumping scheme. Indeed, we stress that the two photon resonance and resulting inversion would not exist except for the influence of the higher lying Jaymes-Cummings levels. This is an important finding for assessing the nonlinear cavity regime. Even in atomic cavity QED systems, signatures of the two-photon resonance with resonant atom-cavity coupling are weak \cite{Rempe:Nature08}.

\begin{figure}[t!]
\centering
\subfigure{
\includegraphics[trim = 2mm 0mm 0mm 0mm, clip, scale=0.6]{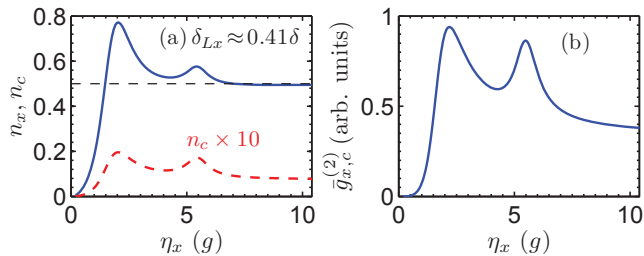}
}
\caption{(Color online)
(a) Exciton (blue) and cavity (red) population, and (b) second-order correlation function versus drive strength, $\eta_x$, for the semiconductor case of Fig.~\ref{fig3}.
\label{fig4}}
\end{figure}

Having demonstrated how to achieve large population inversion through coherent off-resonance excitation in a dissipative semiconductor system, we now consider some common experimental techniques that might detect it. Savage \cite{Savage:PRL88}  suggests that a maximum inversion created by driving at saturation would yield a direct signature in fluorescence. A
method more specifically sensitive to two-photon resonance involves carrying out a photon cross-correlation, i.e.,
measuring the correlation function $\bar g_{x,c}^{(2)}\equiv \braket{ \sigma^+\sigma^- a^\dagger a}_{ss}= \bar g_{c,x}^{(2)}$, which allows one to access information about the quantum statistics and penetrate the pumping cycle that creates inversion. Figures~\ref{fig4}(a-b) display the dependence of the populations and second-order correlation function on the coherent drive strength for a semiconductor system. Two regions of maximum inversion are seen in the populations, with maxima in the cross-correlation function occurring in one to one correspondence.

Finally, it is worth emphasizing that the predicted inversions survive over a very wide range of experimental parameters---e.g., for the semiconductor system, when $\gamma$, $\gamma^\prime$, and $\delta$ all increase by a factor of five, we still obtain peak populations ranging from $0.65-0.85$. In fact, larger detunings can result in larger inversions since this avoids exciting the intermediate states, though, of course, more pump power is required. A more detailed coverage of the parameter dependence,
 including the influence from
 electron-phonon coupling with nonlinear
 coherent driving \cite{phonons1,phonons2}, 
will be described elsewhere.

In conclusion, we have proposed a cavity QED scheme with off-resonance excitation that achieves stationary population inversions
exceeding 0.25 in semiconductor QDs. We also explored the connection to previous measurements in circuit QED and uncovered the limitations of a simplified rate equation model: quantum trajectories clarify the dynamic leading to quantum-noise-induced inversion for dipole couplings ranging over three orders of magnitude in units of cavity loss. More generally, we demonstrated the existence of a cavity-assisted two-photon resonance that significantly enhances the regime of multi-photon cavity QED, in particular for QDs. This can benefit the continued worldwide search for genuine quantum signatures of cavity QED in a number of interesting material systems, along with the development of semiconductor-based single QD lasers~\cite{nomura:Nature10}.

This work was supported by the National Sciences and Engineering Research Council of Canada, the Canadian Foundation for Innovation, and the Marsden Fund of the Royal Society of New Zealand. We thank  S. Reitzenstein, P. Yao and S. S. Shamailov
for useful discussions.


\begin{thebibliography}{10}
\bibitem{Numerical:Note}
To check for numerical convergence in the basis size, we verify that the highest lying moments, $\braket{(a^\dagger)^n a^n}_{ss}$ for an $n$-photon basis, are negligible, and that increasing the basis size does not influence the results.
\end{thebibliography}
\end{document}